\documentclass[twocolumn,preprintnumbers,amsmath,amssymb]{revtex4}
\usepackage{epsfig}
\usepackage{graphicx}% Include figure files
\usepackage{dcolumn}% Align table columns on decimal point
\usepackage{bm}% bold math

\begin{document}

%\preprint{APS/123-QED}

\title{ Mini-review on the temporal resolution of fluorescence imaging systems }% Force line breaks with \\

\author{\bf Partha Pratim Mondal$^{*}$ }
 %\altaffiliation{}%Lines break automatically or can be forced with \\
 %\email{partha@fisica.unige.it}

\affiliation{%
Nanobioimaging Laboratory, Department of Instrumentation and Applied Physics, Indian Institute of Science, Bangalore 560012, India
}%

\date{\today}% It is always \today, today,

\begin{abstract}
Imaging of rapidly occurring events in biology requires high temporal resolution. In this review article, I have tried to investigate an approximate estimate to quantify temporal resolution limit of a fluorescence imaging system. It was realised that, the temporal resolution is essentially determined by the fact that, once excited, almost all ($99.9\%$) excited molecules relax to ground state. We determined the time required for $99.9\%$ of molecules to relax is about $3\tau_p = 3/{(k_f +k_{nr})\log_{10}e}$, where $k_f +k_{nr}$ is the total emission rate. To arrive at this expression we assumed that, the quantum efficiency of the detector is unity, no scattering and there are no other loses. We further discuss few microscopy technique that are capable of high temporal resolution.  These technique includes, multifocal multiphoton microscopy (MMM), multiple excitation spot optical microscopy (MESO) and multiple light-sheet microscopy (MLSM).       
\end{abstract}

\maketitle

%PACS: 87.64.mn, 87.64.kv, 87.85.Pq \\

\section{Introduction}
In the last decade, fluorescence imaging has emerged as one of the key imaging modalities in medicine and biology. The importance of fluorescence techniques have reached a new level with the advent of super-resolution microscopy. Briefly, super-resolution refers to imaging specimens with a resolution better than the classical resolution limit. This limit is popularly known as Abbe's diffraction limit \cite{abbe}. Some of the prominent techniques that are capable of super-resolution are, PLAM \cite{betzig2006}, fPALM \cite{hess2006}, STORM \cite{rust2006}, GSDIM \cite{GSDIM}, STED \cite{sted}, 4pi \cite{hell4pi} and SI \cite{gus2005}. Of-late the field of super-resolution has expanded with the integration of super-resolution with light-sheet microscopy \cite{fra2011}\cite{HessAlby}. Most of these techniques are capable of spatial super-resolution, some at the expense of poor temporal resolution. Specifically, this is true for localization techniques (PALM, fPALM, STORM, GSDIM). On the other hand, fluorescence imaging techniques that are capable of high temporal resolution are, multiphoton multifocal microscopy (MMM), multifocal plane microscopy, multiple excitation spot optical (MESO) microscopy and multiple light-sheet microscopy (MLSM). In the next few sections, we will elaborate on these technique and its working principle. In section II, we derive a simple expression for the temporal resolution of fluorescence imaging system. Section III deals with the description of imaging systems that are capable of high temporal resolution. We conclude the article in section IV. \\

\section{Temporal Resolution Limit in Fluorescence Microscopy}

The most logical way to quantify the temporal resolution can be derived by determining the time required for completing a single excitation-emission cycle of the excited fluorophore. This is based on the fact that, the temporal resolution is ultimately limited by the recycle time of the fluorescent molecule between ground and excited state, provided there is no photobleaching. Consider a simplistic three level system for the fluorophore ($S_0, S_1, T_1$). The following sub-processes occur in a three-level system : Excitation ($S_0 \rightarrow S_1$), Internal Conversion ($S_1 ~(\nu_n \rightarrow \nu_0)$) and Emission ($S_1 \rightarrow S_0$) that occurs in the timescales, $10^{-15} ~s$, $10^{-11} ~s$ and $10^{-9} ~s$ respectively. To simplify the process and to get an approximate estimate of the temporal resolution, one can neglect the time required for excitation and internal conversion as compared to emission process. So, the dominant sub-process that consumes maximum time in an excitation-emission cycle is essentially emission time, and hence the temporal resolution is limited by the time required for all the excited molecules to return to the ground state ($S_0$) from the excited state $(S_1)$. In other words, the temporal resolution is decided by the ensemble of molecules that gets excited at the focal volume. Let us consider that the local density of fluorescent molecules at the focal volume be $\rho_N$ and the focal volume be $V_{psf}$. So, the total number of excited molecules in the focal volume is simply, $\rho_N ~V_{psf}$. During excitation process, the molecules that gets excited to $S_1$ are, $\sigma ~\rho_N ~V_{psf}$, where $\sigma$ is the absorption cross-section. So, the temporal resolution is determined by the the fraction of molecules in the excited state. The limit of temporal resolution is essentially the time required for all the excited molecules (i.e, $\approx 99.9\%$) to return to the ground state. This further assumes the ideal condition that, all the emitted photons are detected. One can take into account many other factors including the detector efficiency for better approximation of temporal resolution limit. \\

In a simplistic three-level system ($S_0, ~S_1, ~T_1$), we assume that, molecules are in excited state $S_1$. There are two major processes that result in deexcitation i.e, fluorescence (with, emission rate, $k_{fl}$) and non-radiative relaxation (with a rate, $k_{nr}$).  The governing equation for determining the excited and triplet state population are given by,
\begin{align}
\begin{cases} 
\frac{\partial}{\partial t} N_{S1}(t) = f_{ill} A ~N_{S0}(t) -(k_f +k_{nr})~ N_{S1}(t)\\ 
\frac{\partial}{\partial t} N_{T1}(t) = k_{nr} N_{S1}(t) 
\end{cases} 
\end{align}
with the additional constraint that, $N_{S0} +N_{S1} +N_{T1} =1$. $f_{ill}\propto h\nu_{ill} k_1$ is the photon flux of the excitation laser. $k_{fl} +k_{nr}$ is total emission rate (including both radiative and non-radiative rate). \\

Since, we are interested in the time required by the molecule to return to the ground state from the excited state, we can start with the initial condition that, the molecules are in excited state $S_1$. Note that, the time required for excitation is orders of magnitude smaller than the emission timescale. Accordingly, one can drop the first term ($ f_{ill} A ~N_{S0}(t)$) on the right-hand side of first rate equation. The resultant rate equation for the excited state population $N_{S1}$ becomes,
\begin{eqnarray}
\frac{d}{dt} N_{S1}(t) = -(k_f +k_{nr})~ N_{S1}(t)
\end{eqnarray}
where $k_{fl} +k_{nr}$ is total emission rate (including both radiative and non-radiative rate). \\

Integration and simplification gives,
\begin{eqnarray*}
\int {\frac{d N_{S1}(t)}{N_{S1}(t)}} = -\int (k_f +k_{nr}) ~dt
\end{eqnarray*}
\begin{eqnarray}
log_{e} (N_{S1}(t)) = -(k_f +k_{nr})t + C
\end{eqnarray}

Now, we divide both the sides by $\log_{10} e$ to convert the solution in $\log_{10}$-scale,
\begin{eqnarray*}
log_{10} (N_{S1}(t)) = -\log_{10}e ~(k_f +k_{nr}) ~t + C_1
\end{eqnarray*}
\begin{eqnarray}
N_{S1}(t) = 10^{-\log_{10}e~(k_{f} + k_{nr})t} \times 10^{C_1} 
\end{eqnarray}
where, $C_1 = C\log_{10}e$. \\ 

Imposing the boundary condition that, $N_{S1}(t)=\sigma ~\rho_N ~V_{psf}$ at $t=0$, we get, $C_1=\log_{10} (\sigma ~\rho_N ~V_{psf})$. Substituting this in (3) gives,
\begin{eqnarray*}
\frac{N_{S1}(t)}{(\sigma ~\rho_N ~V_{psf})} = 10^{-\log_{10}e ~(k_f +k_{nr})  ~t}
\end{eqnarray*}

\begin{eqnarray}
N_{S1}(t) = \sigma ~\rho_N ~V_{psf} ~10^{-t/\tau_p}.
\end{eqnarray}
where, $\tau_p = 1/{(k_f +k_{nr})\log_{10}e}$. \\

Since, the output fluorescence intensity is proportional to the number of molecules in the excited state, so one can rewrite the above equation in terms of output fluorescence,
\begin{eqnarray}
I(t) = I_0 ~10^{-t/\tau_{p}}.
\end{eqnarray}
where, $I_0$ corresponds to the intensity due to $\sigma ~\rho_N ~V_{psf}$ number of molecules in the excitation volume. It may be noted that, the above derivation is similar to the derivation for determining lifetime of fluorescent molecules. \\

Now, we investigate eqn.(5) for few special cases:\\

{\bf{Case. I}}: For $t=\tau_p$, we get, $N(t=\tau_p)=(\sigma ~\rho_N ~V_{psf}) /10$ and hence the corresponding intensity in ideal condition is, $I(t=\tau_p)=I_0 /10$. This indicates the fact that, $\sigma ~\rho_N ~V_{psf}(1-1/10)\%=0.9\sigma ~\rho_N ~V_{psf}$ fraction of molecules (i.e, $90\%$ of total excited molecules) in the focal volume emit before $t=\tau_p$ and $10\%$ emit after $t=\tau_p$. \\

{\bf{Case. II}}: For $t=3\tau_p$, $N(t=3\tau_p)=\sigma ~\rho_N ~V_{psf} /1000$ giving, $I(t=3\tau_p)=I_0 /1000$. This indicates that, $99.9\%$ of total excited molecules emit before $t=3\tau_p$ and so only $0.1\%$ emit after $3\tau_p$. \\

It is to be noted that, the limit on temporal resolution is determined by the time required for nearly cent-percent (i.e, $99.9\%$) molecules to relax to the ground state. As an example, for near cent-percent i.e, $99.9\%$ emission, the molecules require at least $t_{99.9\%} (= 3\tau_p)$ time for completing the cycle ($S_0 \rightarrow S_1 \rightarrow S_0$), assuming all the emitted photons are detected. $t_{99.9\%}$ is the minimum time that the ensemble of molecules requires to prepare itself for the next excitation-emission cycle. This can be considered as the temporal resolution limit and any detector/device that works faster than this time ($t_{99.9\%}=3\tau_p$) would not be able to do better. This restriction is solely due to the molecular excitation-emission cycle of the ensemble of molecules in the focal spot. It may ne noted that, the derivation is simple and similar to the derivation for determining the lifetime of fluorescent molecules. A much more sophisticated estimate can be obtained by considering other factors such as, photobleaching and scattering. \\

\section{Imaging Systems Capable of High Temporal Resolution}

In this section, I will discuss about imaging techniques that are capable of high temporal resolution. Some of the key techniques are: multifocal multiphoton microscopy \cite{bew1998}, multifocal plane microscopy \cite{ram2012}, multiplane microscopy \cite{dalgarno2010}, multiple excitation spot microscopy \cite{partha2011} and multiple light-sheet microscopy \cite{parthaPLOS1}. Here we will focus on some of these techniques.

\subsection{Multifocal Multiphoton Microscopy}

Multifocal multiphoton microscopy is probably the first fluorescence imaging technique that brings in the concept of high temporal resolution. Multiphoton excitation takes place in the focal region where the high photon flux makes possible the simultaneous interaction of two photons with the target molecules. With the availability of powerful lasers (Ti-Saffire laser) of average power of $\approx 1 ~W/cm^2$ and peak power in few hundred Gigawatts, it becomes attractive to think of ways to accelerate the imaging process by splitting the beam so as to generate several focal spots. This facilitates large area parallel scanning of the specimen. \\

In ideal situation, the signal from the focus of a 2-photon excitation (2PE) microscopy system is directly proportional to the $2~nd$-order of average laser power $P_a$, and inversely proportional to pulse width ($\tau$) and repetition rate ($f$) i.e,
\begin{eqnarray}
I_{nPE}=\sigma_2 ~\frac{P_a^2}{(\tau ~f) }
\end{eqnarray}
where, $\sigma_2$ is the n-photon cross-section. \\

In case of MMM system, the laser beam is split into M-beams so the average power gets scaled by $M$ i.e, the power for each beam in MMM system is, $P_{MMM} = P_a /M$. So the output signal emerging from the focus for the overall MMM system of M-beams is given by,
\begin{eqnarray}
I_{MMM,~2PE}=\sigma_2 ~\frac{P_{MMM}^2}{\tau ~f~M }
\end{eqnarray}
where, $\sigma_2$ is the 2-photon absorption cross-section. \\

\begin{figure}
\includegraphics[height=2.2in,angle=0]{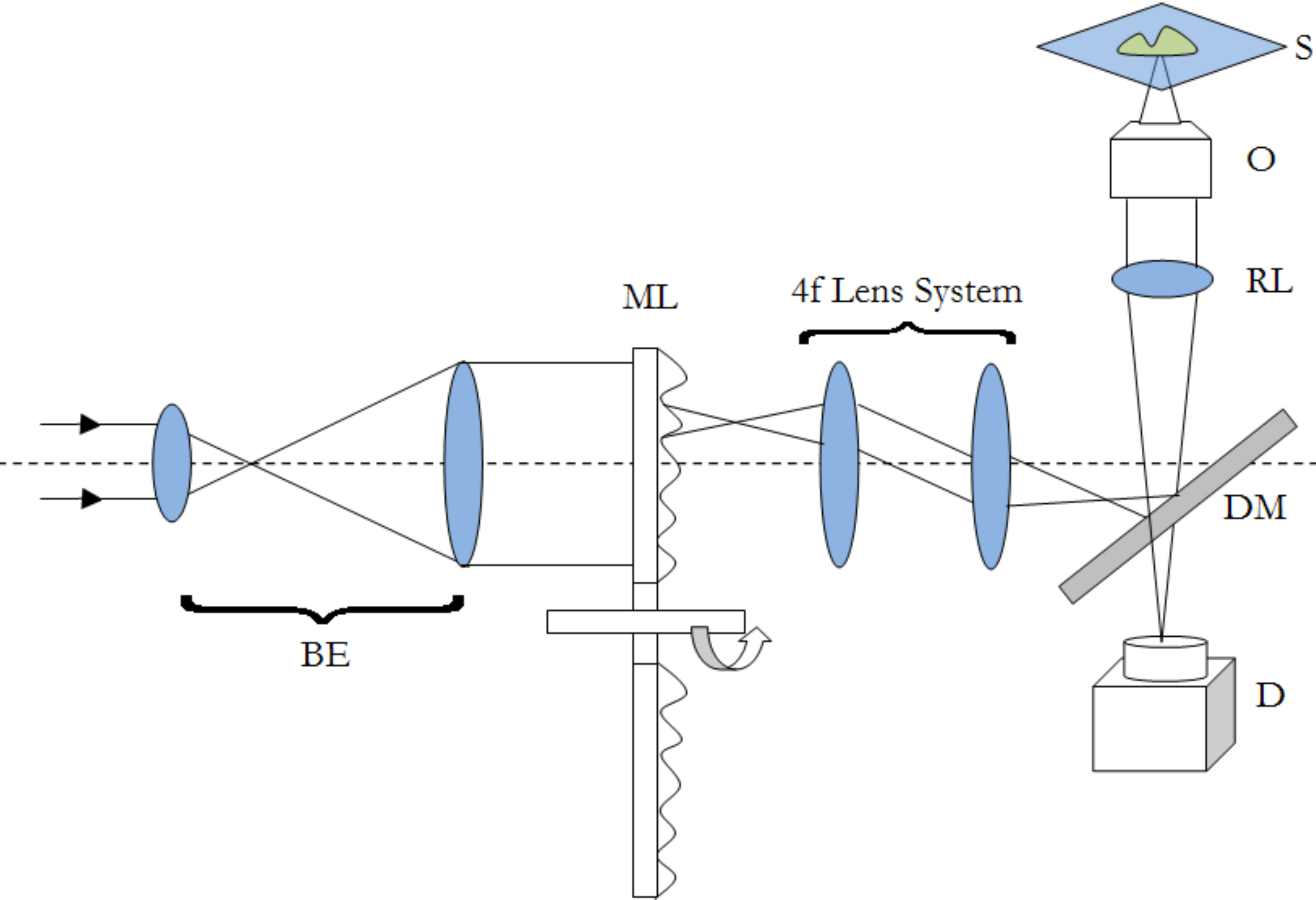}
\caption{\small{ Schematic diagram of a typical MMM imaging system.}}
\end{figure}

%\begin{figure}
%\centerline{\psfig{figure=MMM_Fig1.eps,height=6cm,angle=0} }
%\caption{\small{ Schematic diagram of a typical MMM imaging system.}}
%\end{figure}

A typical realization of MMM-system is schematically shown in Fig. 1. There are two parts: first, the microlens array and second, the high resolution inverted fluorescence microscope. The laser beam is appropriately expanded using a beam-expander (BE) to fill the microlens array. This results in the generation of many tiny focal spots which are directed to the dichoric mirror (DM) using  a 4f lens system. The rays are subsequently reach the back-aperture of the objective lens through relay lens (RL). For simplicity, ray diagram for only one of the lens is shown in Fig.1. Similarly, an array of discrete focus points are created in the focal plane by other micro-lenses. Subsequently, the fluorescence from each of the points are collected by the same objective lens (O) and focused to the detector (D). The dichroic mirror filters out the incident light and allows the Stoke-shifted fluorescence light to pass-through. This system results in imaging many points on the plane in a single go and the disk can be rotated (see Fig.1) to collect data from other planes. In this way fast data-collection can be achieved. As an example, xy-images of live boar-sperm cells are obtained within 30 ms using MMM system has been demonstrated \cite{bew1998}. The head and tail of the cells are respectively labeled with Hoechst 33342 and fluorescein. A stack of 191 images were reportadely obtained within 6s that help in monitoring of the movements of the cells in the aqueous medium \cite{ptso2007}. A recent version of MMM system indicate a frame rate of $>600 Hz$. Using this system, contraction of cardiac myocytes (labeled with Fluo3 dye) were obtained at an incredible fast rate of 640 Hz \cite{ptso2007}.   \\  

Noted that, there are many more realizations of MMM system \cite{pawleyBook}. Moreover, advanced variants of MMM system include, fluorescence lifetime imaging and second harmonic generation that are capable of generation multi-focal lifetime and second-harmonic images much faster than the classical techniques. While MMM system has many advantages they have limitations too. The most prominent one is the crosstalk arising from the overlap of the fields from multiple foci. Another limitation is the inability to obtain reliable image of scattering specimen. \\

\subsection{Multiple Excitation-Spot Optical Microscopy}

Multiple excitation spot optical microscopy (MESO) is an imaging technique that enables simultaneous visualization of multiple specimen planes. Unlike other techniques, this has the inbuilt potential for enabling multi-particle tracking. This feature makes it a unique tool for simultaneous monitoring and tracking many particles for understanding complex biological processes. This technique was first proposed by us \cite{partha2009}\cite{partha2011}.  \\

The schematic diagram describing a MESO imaging system is shown in Fig.8. The illumination consists of objective lens $O_1$ and $O_2$ arranged in a $4\pi$-geometry. Spatial filtering technique is employed in which both the objective lens are aperture engineered. Spatial filter is purposefully designed to produce an annular/ring illumination at the back of the objective lens. Since, the objective lens approximately perform Fourier transform, this give rise to Bessel-like beam when illuminated with a ring-illumination pattern. A similar counter-propagating phased-matched Bessel-like beam is generated by the aperture-engineered objective $O_2$. The super-position of both the beams at the focus give rise to interference pattern producing an array of nanodot pattern. Orthogonal detection is employed to detect fluorescence from individual nanodots. This is achieved by placing a detection objective $O_3$ at $90^o$ to the illumination arm (see, Fig.8). The detection object is moved along the optical z-axis to collect fluorescence from individual nanodots. \\

The overall system PSF of the MESO imaging system id determined by the illumination and detection PSFs. Note that, the illumination is aperture-engineered in a $4\pi$-geometry. The $x$-, $y$- and $z$- components of the electric field for randomly polarized light illumination in Cartesian coordinate system are given by \cite{wolf1959}\cite{biovin1965}\cite{partha2011}, 
\begin{eqnarray}
\Biggl{[}\begin{array}{ll}  E_x \\ E_y \\ E_z \end{array}\Biggr{]}= \Biggl{[}\begin{array}{ll} -iA(I_0 + I_{2} \cos(2\phi))\\ -iAI_{2} \sin(2\phi)\\ -2A I_{1} \sin(\phi) \end{array}\Biggr{]}
\end{eqnarray}
where $A$ is the proportionality constant representing the amplitude of the incident electric field. \\

For randomly-oriented dipoles, the excitation PSF of the MESO system is,
\begin{eqnarray*}
h_{ill}(x,y,z)=|\bar{E}(x,y,z) +\bar{E}(x,y, -z)|^2 
\end{eqnarray*}
\begin{eqnarray}
= |Re{\tilde{I}_{0}}|^2 + 2|Re{\tilde{I}_{1}}|^{2} + |Re{\tilde{I}_{2}}|^{2} 
\end{eqnarray}
where, the modified diffraction integral over the aperture angle $\alpha_{ill}$ are given by,
\begin{eqnarray*} 
\tilde{I}_{0,1,2}= \int_{\theta=0}^{\alpha_{ill}} B(\ast) G_{0,1,2}(\ast) {\cos^{1/2}{\theta}} ~e^{ i(\frac{u\cos{\theta}}{\sin^2 {\alpha_{ill}}})} d\theta 
\end{eqnarray*} 
and, 
\begin{eqnarray*}
\Biggl{[}\begin{array}{ll} G_0(\ast)\\  G_1(\ast)\\ G_2(\ast) \end{array} \Biggr{]} = \Biggl{[}\begin{array}{ll} \sin{\theta} (1+\cos{\theta}) J_0(\frac{v\sin{\theta}}{sin{\alpha}}) \\ ~\sin^2{\theta} J_1(\frac{v\sin{\theta}}{sin{\alpha}}) \\ ~\sin{\theta}(1-\cos{\theta}) J_2(\frac{v\sin{\theta}}{sin{\alpha}}) \end{array} \Biggr{]}
\end{eqnarray*}

The spatial filter is characterized by the optical mask with transmission function, 
\begin{eqnarray}
B(\ast)=H[\theta-\theta_1]-H[\theta-\alpha]
\end{eqnarray} 
where, $H(*)$ is the Heaviside function. The cutoff angle is, $\theta_1$, for which the transmission window is $\Delta\theta=5^o$. $u=\frac{2\pi}{\lambda}{z\sin^2{\alpha}}$ and $v={\frac{2\pi}{\lambda} {x^2 +y^2}^{1/2}\sin\alpha}$ are the longitudinal and transverse optical coordinates, respectively \cite{wolf1959}\cite{biovin1965}, and $\alpha_{ill}$ is the illumination semi-aperture angle. \\ 

The technique employs a theta detection scheme, in which the detection is carried out in the orthogonal plane, as shown in Figure 1. The orthogonal plane is represented by the following transformation : ($x^{\prime}=-z, ~y^{\prime}=y, ~z^{\prime}=x$). This detection scheme has the advantage of high resolution at long working distance. The isotropic emission model is assumed with randomly polarized light excitation. The components of the detection PSF are, 
\begin{eqnarray}
h_{det}(x^{\prime}, y^{\prime}, z^{\prime} )=|\bar{E}_{x^{\prime}}|^2 +|\bar{E}_{y^{\prime}}|^2 + |\bar{E}_{z^{\prime}}|^2 
\end{eqnarray}

where the integration $I_{0,1,2}= \int_{\theta=0}^{\alpha_{det}} (...) d\theta$ on the aperture-free objective ($O_3$) is carried over the detection semi-aperture angle $\alpha_{det}$. \\

Overall, the system PSF of the proposed imaging system is given by, 
\begin{eqnarray}
h_{sys}(x,y,z) = h_{ill}(x,y,z) \times h_{det} (-z,y,x).
\end{eqnarray}

The illumination, detection and system PSF for the proposed $MESO$ system are shown in Fig.9 for $\alpha_{ill}=45^o$ and $\alpha_{det}=45^o$. One can immediately note that, the axial resolution of the proposed system is improved by a factor of approximately $4$ over the classical resolution limit, while there is no improvement in the lateral resolution because the perpendicular detection occurs along the lateral axis of the illumination objective. Study show that, the axial and lateral resolutions are approximately, $120~nm$ and $180~nm$, respectively \cite{partha2011}. In principle, the fluorescence from all of the nanodots can be recorded by scanning the objective $O_3$. In principle, simultaneous detection from all nanodots can be obtained by using appropriate optical elements, such as, a distorted diffraction grating in the detection path of the imaging system \cite{dalgarno2010}. Such a system has the capability for near-simultaneous in-depth multilayer imaging of biological specimens, thereby increasing the temporal resolution of the imaging system. For example, to scan $M$ slices of an $N\times N$ image, a laser scanning system requires $M\times N^2$ scan-points and a total time of $(M\times N^2)\Delta t$. MESO system with $M$ axial dots, allows in-principle the data acquisition rate to be increased by a factor of $M$, thereby increasing the temporal resolution by a factor of $M \Delta t$.  \\ 
   
\begin{figure}
\includegraphics[height=2.1in,angle=0]{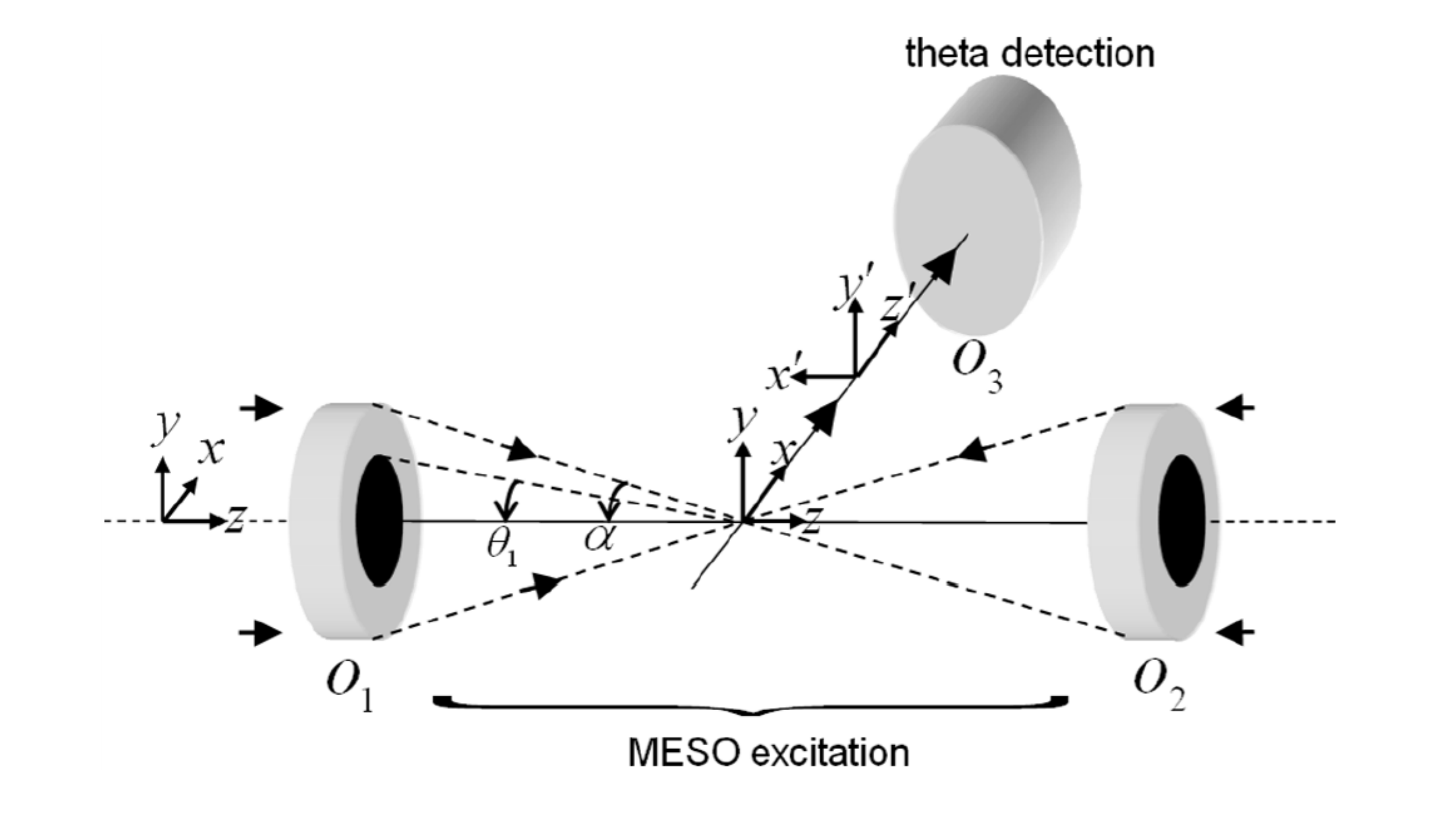}
\caption{\small{ Schematic diagram describing the optical configuration of MESO microscopy.}}
\end{figure}

\begin{figure}
\includegraphics[height=3.0in,angle=0]{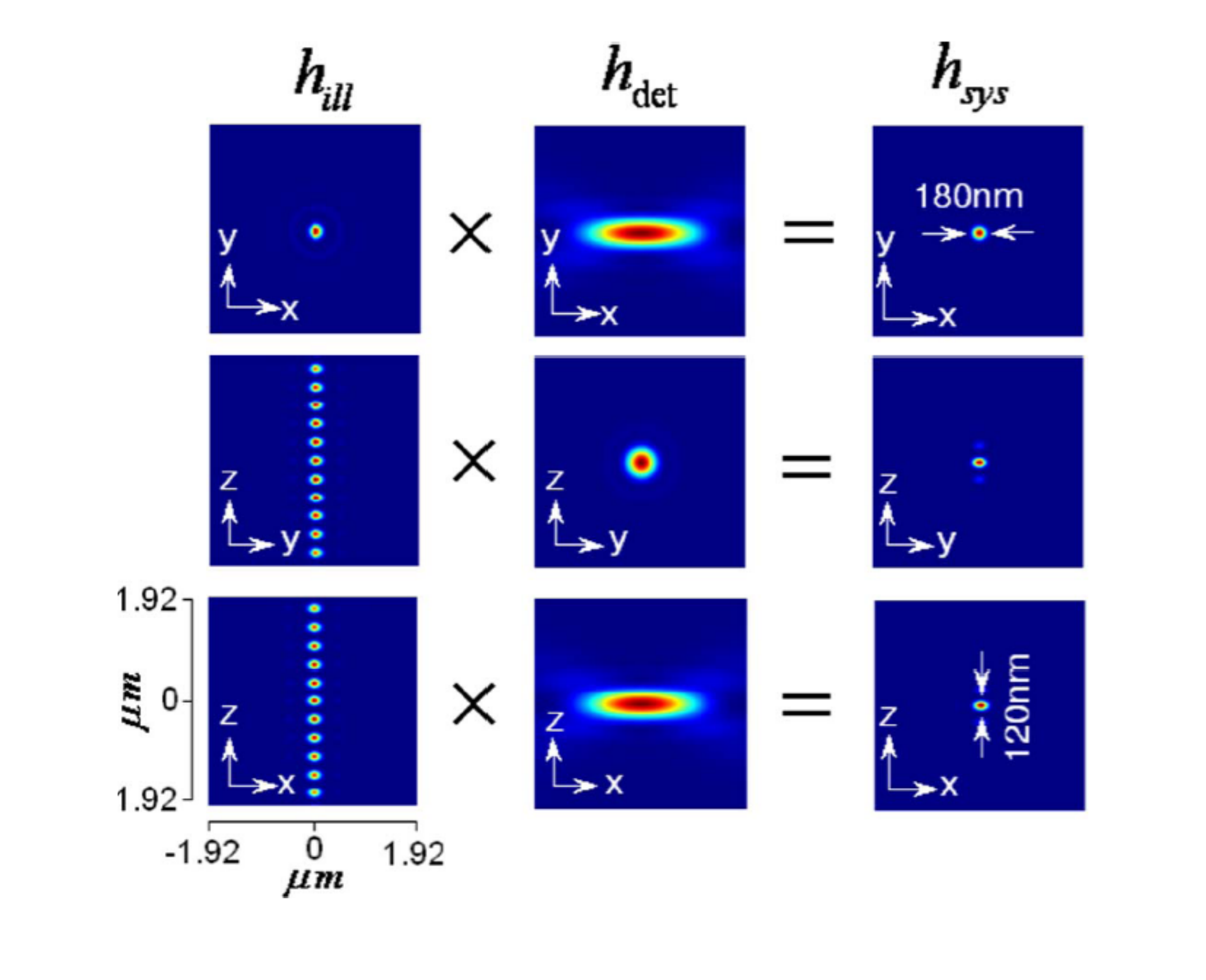}
\caption{\small{Illumination, detection and system PSF for MESO system \cite{partha2011}. }}
\end{figure}

\begin{figure*}
\includegraphics[height=4.5in,angle=0]{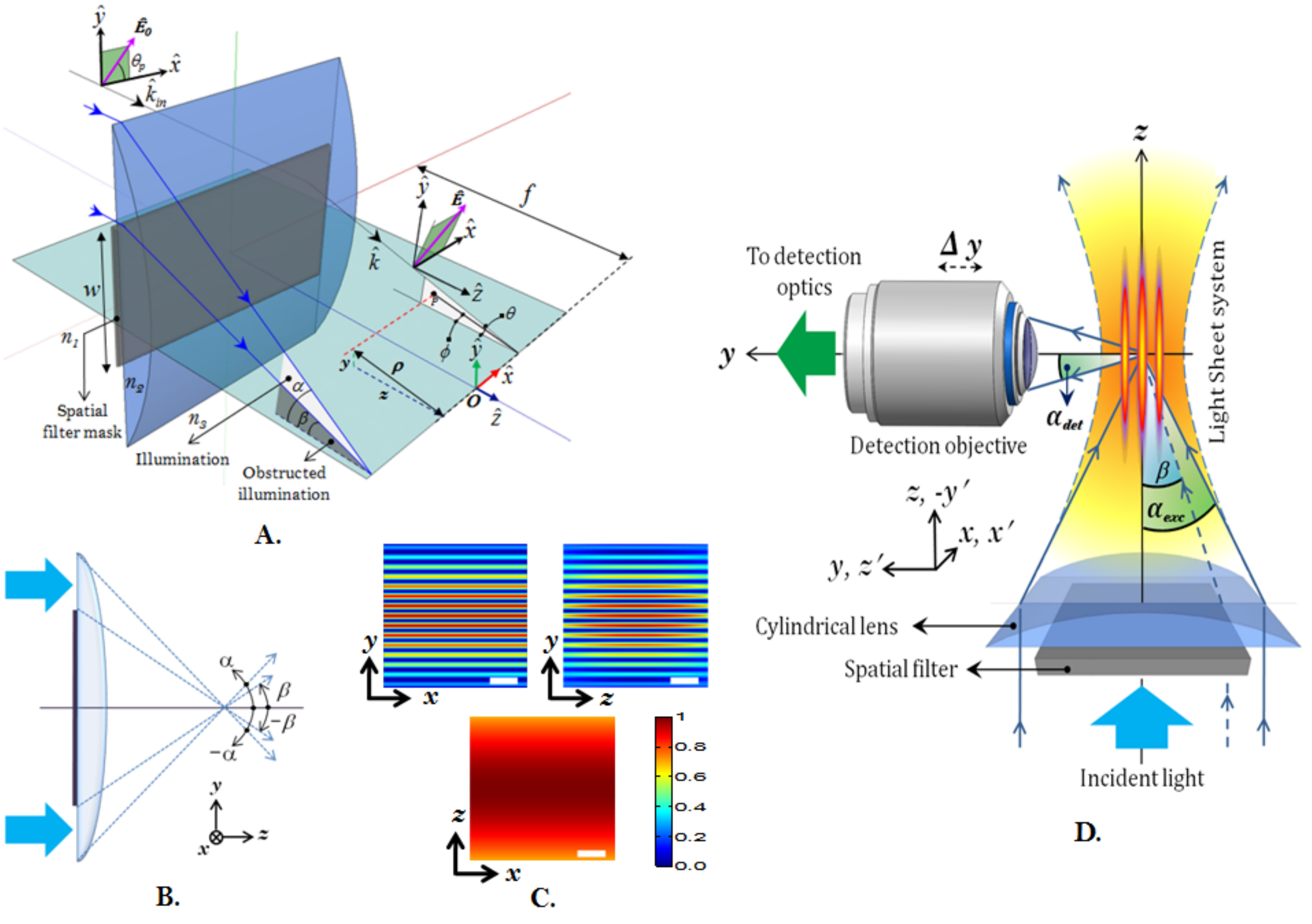}
\caption{\small{(A) Schematic diagram of the excitation part describing all the imaging parameters \cite{parthaPLOS1}. (B) Spatial filter. (C) $XY$, $YZ$, $XZ$ planes of excitation PSF (imaging parameters: $\lambda_{exc} = 532nm$, $\alpha_{exc} = 30^o$, $\beta = 25^o$ ), (D)  Schematic diagram of the complete imaging system with orthogonal detection. Scale-bar$= 1.5 \mu m$. }}
\end{figure*}

\begin{figure*}
\includegraphics[height=4.5in,angle=0]{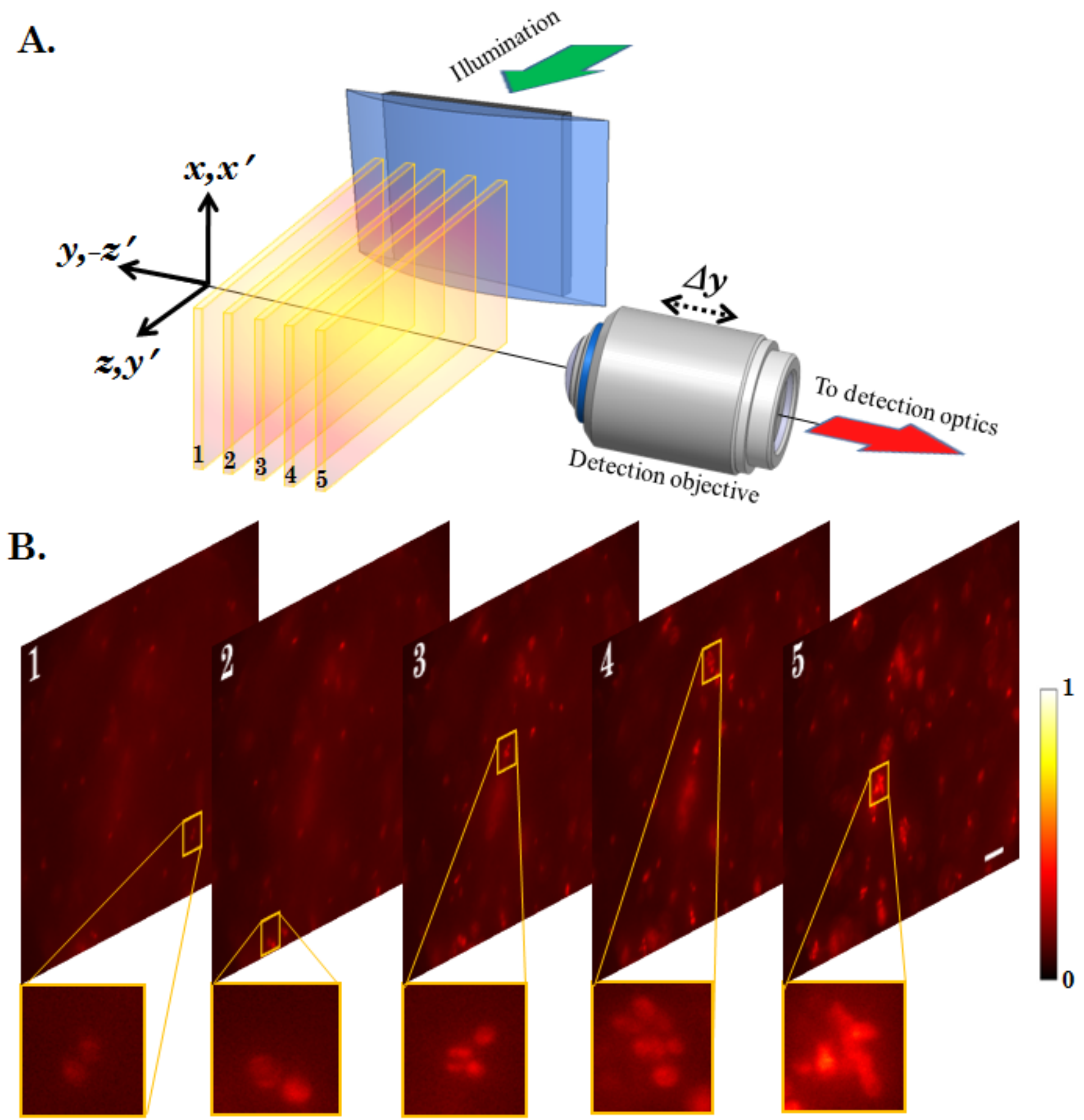}
\caption{\small{(A) Schematic diagram showing the optical setup of the MLSM system \cite{parthaPLOS1}. (B) 2D images obtained from different planes of the specimen probed by multiple light-sheet pattern. Scalebar : $25 \mu m$. }}
\end{figure*}

\subsection{Multiple Light Sheet Microscopy}

This microscopy technique involves the generation of multiple light-sheets and imaging. Spatial filtering technique is employed in a SPIM microscopy setup to generate multiple light sheets. The technique is robust and paves the way for volume imaging in fluorescence microscopy \cite{parthaPLOS1}. \\

The system PSF is solely determined by the structure of the spatial filter employed in a cylindrical lens system. The incident plane wave is subjected to rectangular spatial filter that allows the light to pass through the periphery (see, Fig.11A). The structured wavefront thus created was incident at the back aperture of the cylindrical lens. Since the cylindrical lens performs 1D fourier transform, this results in multiple light-sheets at and off the focus. The intensity of the light-sheets differ in intensity but can be tailored by adjusting the parameters of spatial filter. The resulting field distribution can be calculated at and near the focus. In cylindrical coordinates ($\rho,\phi,z$) with primary axis of the cylindrical lens along $x-$direction, the electric field components at the focus for a linearly polarized light illumination (polarization angle $\theta_p$ with $x$-axis) with profile $\left. E_{in}(\theta)\right.$ is given by \cite{parthaAIP3},     
\begin{align}
		&\!\!\!\!\!\left[ \!\!\!\!\begin{array}{c} \left. E_x(\rho,\phi)\right. \\ \left. E_y(\rho,\phi)\right. \\ \left. E_z(\rho,\phi)\right. \end{array}\!\!\!\! \right]\!\!=\!\!A \!\!\!\int\limits_{-\alpha}^{\alpha}\!\!\left. E_{in}(\theta)\right. {\left[\!\!\!\!\begin{array}{c}\cos\theta_p \\ 
		\sin \theta_p \cos\theta \\ \sin \theta_p \sin\theta \end{array} \!\!\!\!\right]}\!\! \cos^{\frac{1}{2}}\!\theta e^{[i\rho k\{\cos(\theta-\phi)\}]} \text d\theta\!\! \label{eq1}
\end{align}
where, $\alpha$ is the semi-aperture angle of the lens defined by it's numerical aperture, $k$ is the wavenumber in the image space and $A=\sqrt{\frac{fk}{2\pi}} e^{-ifk} e^{i\pi /4}  \sqrt{\frac{n_1}{n_3}}$. The terms $f$,$n_1$ and $n_3$ denote the focal length, refractive index of object and image space respectively while the radial distance from the $x-$axis and the polar inclination are denoted by, $\rho=\sqrt{y^2+z^2}$ and $\phi=\tan^{-1}(y/z)$ respectively. By introducing an amplitude transmission function $T(\theta), |\theta| \leq \alpha$ (that of spatial filter), Eq.(\ref{eq1}) modifies to,
\begin{align}
		&\!\!\!\!\!\left[ \!\!\!\!\begin{array}{c} E_x\\ E_y\\ E_z\end{array}\!\!\!\! \right]\!\!=\!\!A \!\!\!\int\limits_{-\alpha}^{\alpha}\!\! \left. E_{in}(\theta)\right. T(\theta){\left[\!\!\!\!\begin{array}{c}\cos\theta_p \\ 
		\sin \theta_p \cos\theta \\ \sin \theta_p \sin\theta \end{array} \!\!\!\!\right]}\!\! \sqrt{\cos\theta} e^{[i\rho k\{\cos(\theta-\phi)\}]} \text d\theta\!\! \label{eq31}
\end{align}  

The corresponding transmission function $T(\theta)$ of the spatial filter that is suitable for generating multiple light-sheets is given by,
\begin{align}
T(\theta)= \begin{cases} 1, & \mbox{if } \beta < |\theta| \leq \alpha\\ 0, & \mbox{if } |\theta|\leq\beta \end{cases} \label{eq34}
\end{align}

The schematic diagram of MLSM microcopy system is shown in Fig. 11D. The system can be broadly split into two independent optical configurations : excitation sub-system and detection sub-system. Orthogonal detection system is employed which embodies many advantage over the existing detection techniques. Light of wavelength $\lambda_{exc} = 532~nm$ is allowed to pass through the spatial filter (transmission function $T(\theta)$),that generate structured wavefront, $S(\theta)=A(\theta)T(\theta)$, where, $A(\theta)=1$ for plane wave. The excitation sub-system and the spatial filter along with the imaging parameters are shown in Fig.11A and Fig.11B respectively. The illumination PSF of the proposed imaging modality is obtained using eqn.(16) and the result is shown in Fig. 11C. A central light-sheet along with light-sheets on either side is clearly evident. Although weak the intensity of off-focus light-sheets can be boosted by tuning the stop angle ($\beta$) of the spatial filter. This facilitates simultaneous excitation of multiple specimen layers. The intensity of central and few nearby sidelobes are fairly intense but falls of gradually thereafter. This can be explained based on the 1D Fourier transform (performed by the cylindrical lens) of a rectangular window function (spatial filter)which is a $Sinc$ function (intensity distribution at focus). This further evidences the fact that, one can reliably scan $5-7$ layers of the specimen in a single shot. This is a step closer to volume imaging.The complete MLSM system that comprises both illumination and detection sub-systems is shown in Fig.11D. The fluorescence light emanating from the specimen is collected by the detection objective placed orthogonal to the illumination arm. The light is subsequently filtered to remove the scattered light and focused on the CCD camera. \\

We used multiple light-sheet illumination and a dedicated orthogonal detection system to realize a complete imaging system. This is accomplished by translating the detection arm in order to focus on a specific light-sheet in a selective manner. The results on a Agarose gel sample with suspended yeast cells (coated with NHS-Rhodamine tagged polymer) are shown in Fig.13 \cite{parthaPLOS1} Translation along $+z^{\prime}$ axis reveal a series of in-focus and out-of-focus image planes. The in-focus planes are observed as a result of intersection of a particular light-sheet with the detection PSF. Note that, we also observed out-of--focus fluorescence from other specimen planes. The background is created by other light-sheets separated by some distance along the $y-$ axis which appear as defocused objects. Better detection system along with high NA objective (that shrinks the detection PSF) can be employed to reduce the contribution from out-of-focus light. Overall the proposed system may enable simultaneous monitoring of different organ growth during the developmental process and expedite volume imaging. \\

\section{Conclusions}
In this review article, we derived a simple and approximate expression to determine the temporal resolution. It was realized that, the temporal resolution depends on the recycle time (emission-excitation cycle) of $99.9\%$ of the excited fluorophore at the focal volume. We further discuss some of the prominent optical microscopy system that are capable of improving the temporal resolution. These techniques use multiple excitation spots/planes that brings in simultaneous imaging of multiple planes. Surprisingly, temporal resolution does not depend on the imaging speed (frames/sec) of the detector. \\

\end{document}